\documentclass[a4paper]{jpconf}
\usepackage{graphicx}
\usepackage{amsmath}
\newcommand{\beq}{\begin{equation}}
\newcommand{\eeq}{\end{equation}}
  \newcommand{\beql}[1]{\begin{equation}\label{eq:#1}}
  \newcommand{\beqa}{\begin{eqnarray}}
  \newcommand{\eeqa}{\end{eqnarray}}
  \newcommand{\beqas}{\begin{eqnarray*}}
  \newcommand{\eeqas}{\end{eqnarray*}}

  \newcommand*{\R}{\mathbf{R}}

  \newcommand*{\bM}{\mathbf{M}}
  \newcommand*{\bx}{\mathbf{x}}
  
  \newcommand*{\cB}{\mathcal{B}}
  \newcommand*{\cC}{\mathcal{C}}

  \newcommand*{\cS}{\mathcal{S}}
  \newcommand*{\da}{\dagger}
 \newcommand*{\de}{\delta}
 \newcommand*{\ta}{\tau}                                      
  
  \newcommand*{\Ga}{\Gamma}                                          
  
\newtheorem{Theorem}{Theorem}

\newcommand{\kps}{\ket{\psi}}
\newcommand{\kph}{\ket{\phi}}
\newcommand{\kxi}{\ket{\xi}}

  \newcommand*{\dd}{d}
 \newcommand*{\bP}{\mathbf{P}}
  \newcommand*{\bS}{\mathbf{S}}
  
  \newcommand*{\cH}{\mathcal{H}}
  \newcommand*{\cK}{\mathcal{K}}
  
  \newcommand*{\al}{\alpha}
  \newcommand*{\be}{\beta} 
  \newcommand*{\ep}{\epsilon}
   \newcommand*{\eq}[1]{(\ref{eq:#1})}
  \newcommand*{\et}{\eta}
  
  \newcommand*{\nn}{\nonumber}
  \newcommand*{\om}{\omega}
  \newcommand*{\ps}{\psi} 
 \newcommand*{\rh}{\rho}
  \newcommand*{\si}{\sigma} 
  \newcommand*{\ve}{\varepsilon}
  \newcommand*{\De}{\Delta}                                          
  \newcommand*{\Eq}[1]{Eq.~(\ref{eq:#1})}
  \newcommand*{\Om}{\Omega}
  \newcommand*{\Th}{\Theta}                                          
\newcommand*{\bra}[1]{\langle#1|}
\newcommand*{\ket}[1]{|#1\rangle}
\newcommand*{\braket}[1]{\langle#1\rangle}
\newcommand*{\bracket}[1]{\langle#1\rangle}
\newcommand*{\ketbra}[1]{\ket{#1}\bra{#1}}

\newcommand{\x}{Q}
\newcommand{\px}{P}
\newcommand{\y}{\overline{Q}}
\newcommand{\py}{\overline{P}}

\renewcommand{\o}{\om}
\newcommand{\Det}{\Delta t}

\renewcommand{\Theta}{X}
\renewcommand{\Th}{X}
\renewcommand{\theta}{x}
\renewcommand{\Om}{Y}
\renewcommand{\om}{y}
\renewcommand{\ep}{\varepsilon}

\begin{document}
\title{Heisenberg's uncertainty relation: Violation and reformulation}

\author{Masanao Ozawa}

\address{Graduate School of Information Science,
Nagoya University, Chikusa-ku, Nagoya, 464-8601, Japan}

\ead{ozawa@is.nagoya-u.ac.jp}

\begin{abstract}
The uncertainty relation formulated by Heisenberg in 1927 describes a
trade-off between the error of a measurement of one
observable and the disturbance caused on another complementary
observable so that their product should be no less than a limit set by
Planck's constant.  In 1980, Braginsky, Vorontsov, and Thorne claimed
that this relation leads to a sensitivity limit for gravitational wave
detectors.  However, in 1988 a model of position measurement was 
constructed that breaks both this limit and Heisenberg's relation.
Here, we discuss the problems as to how we reformulate Heisenberg's
relation to be universally valid and how we experimentally quantify
the error and the disturbance to refute the old relation and to confirm
the new relation.
\end{abstract}

\section{Heisenberg's EDR}

The discovery of quantum mechanics introduced 
non-commutativity in algebraic calculus of observables;
the {\em canonical commutation relation} (CCR)
\beql{CCR}
[Q,P]=i\hbar
\eeq
is required to hold between a coordinate $Q$ of a particle and its momentum $P$,
where the commutator $[Q,P]$ is defined by  $[Q,P]=QP-PQ$.
In 1927, Heisenberg proposed 
an operational meaning of the non-commutativity:
``the more precisely the 
position is determined,  the less precisely the momentum is known, and 
conversely'' \cite[p.~64]{Hei27E}. @  

By the famous $\gamma$ ray microscope 
thought experiment he derived the relation
\beql{Hei27E}
\ep(Q)\et(P)\ge\frac{\hbar}{2},
\eeq
where $\ep(Q)$ is the ``mean error'' of a position measurement
and $\et(P)$ is the thereby caused ``discontinuous change'' in the momentum $P$:
\begin{quote}
Let $\ep(Q)$ be the precision with which
the value $Q$ is known ($\ep(Q)$ is, say, the mean error of $Q$), therefore here the wavelength
of the light. Let $\et(P)$ be the precision with which the value $P$ is determinable; that is,
here, the discontinuous change of $P$ in the Compton effect \cite[p.~64]{Hei27E}. 
\end{quote}
Here, ``mean error'' is naturally understood to be ``root-mean-square (rms) error ''
as introduced by Gauss \cite{Gau95}, and ``discontinuous change'' is often called ``mean disturbance.''
Heisenberg  claimed that \Eq{Hei27E} is a ``straightforward mathematical consequence'' of \Eq{CCR}
\cite[p.~65]{Hei27E} and
gave its mathematical justification \cite[p.~69]{Hei27E}.

\section{Heisenberg's derivation}

In his mathematical justification of \Eq{Hei27E},  
Heisenberg firstly
derived the relation
\beql{Hei27ESD}
\sigma(Q)\sigma(P)=\frac{\hbar}{2}
\eeq
for the standard deviations $\si(Q)$ and $\si(P)$
of the position $Q$ and the momentum $P$
in the state described by a Gaussian wave function \cite[p.~69]{Hei27E},
which Kennard \cite{Ken27} subsequently  generalized as the relation
\beql{Ken27}
\sigma(Q)\sigma(P)\geq\frac{\hbar}{2}
\eeq
for arbitrary wave functions.
Here,
the standard deviation is defined for any observable $A$ 
by $\si(A)^2=\bracket{A^2}-\bracket{A}^2$,
where $\bracket{\cdots}$ stands for the mean value in a given state.
Note that in Ref. \cite[p.~69]{Hei27E} Heisenberg actually derived the relation 
\beq 
\tilde{\si}(Q)\tilde{\si}(P)={\hbar}
\eeq
for $\tilde{\si}(Q)=\sqrt{2}\si(Q)$ and $\tilde{\si}(P)=\sqrt{2}\si(P)$
in Gaussian wave functions, and 
Kennard \cite{Ken27} actually derived the relation 
\beq
\tilde{\si}(Q)\tilde{\si}(P)\ge \hbar
\eeq
that generalizes Heisenberg's relation to arbitrary wave functions.

Heisenberg secondly applied \Eq{Ken27} to the state just after the measurement
assuming: 
\begin{enumerate}
\item[\small (H1)] {\em Any measurement with rms error $\ep(A)$
of an observable $A$ leaves the state satisfying $\si(A)\le\ep(A)$.}
\item[\small (H2)] {\em If an observable $A$ can be measured with the rms error 
$\ep(A)$ and the rms disturbance $\et(B)$ of another observable $B$,
then $A$ and $B$ can be jointly measured with the rms errors
$\ep(A)$ and $\ep(B)=\et(B)$, respectively.  }
\end{enumerate}
Then, it can be easily seen that 
\Eq{Hei27E} can be derived from \Eq{Ken27} under assumptions
(H1) and (H2).
In fact, if $Q$ can be measured with $\ep(Q)=\al$ and $\et(P)=\be$, then by (H2) 
$Q$ and $P$ can be measured jointly with $\ep(Q)=\al$ and 
$\ep(P)=\be$, so that by 
(H1) the state after the joint measurement satisfies $\si(Q)\le\al$
and $\si(P)\le\be$, and hence \Eq{Ken27} concludes \Eq{Hei27E}.

\section{Heisenberg's unsupported assumption}
Assumption (H2) is considered to hold in general; see Ref.~\cite{03UVR} for a detailed discussion.
However, assumption (H1) is not, whereas
Heisenberg's contemporaries, including von Neumann, supported 
assumption (H1):
\begin{quote}
We are then to show that if $Q, P$ are two canonically conjugate quantities, 
and a system is in a state in which the value of $Q$ can be given with the
accuracy $\ep [=\si(Q)]$ (i.e., by a $Q$ measurement with an error
range $\ep [=\ep(Q)]$), then $P$ can be known with no greater accuracy
than $\et [=\si(P)]=\hbar/(2\ep)$. Or: a measurement of $Q$ 
with the accuracy $\ep [=\ep(Q)]$ must bring about an indeterminancy 
$\et [=\et(P)]=\hbar/(2\ep)$ in the value of $P$ \cite[pp.~238--239]{vN55}.
(Terms in $[\ldots]$ are supplemented by the present author.)
\end{quote}
In those days the repeatability hypothesis was considered a natural requirement 
for all the precise measurements of an observable $A$ and (H1) is considered as
a natural generalization to arbitrary approximate measurements of $A$.
Here,  the {\em repeatability hypothesis} is formulated as follows.
\begin{itemize}
\item[{\small (RH)}] {\em If an observable $A$ is measured twice in succession in a system, 
then we get the same value each time} \cite[pp.~335]{vN55}.
\end{itemize}
Under (RH), any precise measurement of $A$ with $\ep(A)=0$ changes the state 
to be an eigenstate of the measured observable $A$, which satisfies $\si(A)=0$.
However, in the light of modern theory of quantum measurement, 
(RH) has been abandoned as proposed by Davies and Lewis \cite{DL70}:
\begin{quote}
One of the crucial notions is that of repeatability which we show is implicitly
assumed in most of the axiomatic treatments of quantum mechanics, but whose
abandonment leads to a much more flexible approach to measurement theory
\cite [p.239]{DL70}.
\end{quote}
In fact, in Ref.~\cite{84QC} we have mathematically characterized all the physically possible 
quantum measurements, shown that (RH) is no longer universally valid, and 
even more  that no precise measurements of continuous observables satisfy (RH). 
Thus, (H1) does not hold even in the case where $\ep(A)=0$.

Therefore, in the light of the modern theory of quantum measurement,
assumption (H1) cannot be  accepted, so that  \Eq{Hei27E} cannot be considered as
an immediate consequence of \Eq{Ken27}, although their meanings have often 
been confused even in standard text books \cite{vN55,Boh51,Mes59a,Sch68}.
As above, the original justification of \Eq{Hei27E} was limited, 
but its universal validity was not been refuted in theory until 1980's.

\section{Von Neumann's model of position measurement}

Until 1980's only solvable model of position measurement had been given 
by von Neumann \cite{vN55}.
In what follows, we discuss the von Neumann model and show that this 
long-standing standard model satisfies the Heisenberg error-disturbance relation (EDR) \eq{Hei27E}.
Thus, the model analysis of position measurement did not lead to refuting the Heisenberg 
EDR but rather enforced the belief that good position measurements 
satisfy the Heisenberg EDR \eq{Hei27E}.

Consider a one-dimensional mass, called an {\em object},
with position $\x$ and momentum $\px$, described by a Hilbert space $\cH$.
The measurement of $\x$ is carried out by a coupling between the object $\bS$ and a probe $\bP$
from time $t=0$ to $t=\Det$.
The probe $\bP$ is another one-dimensional mass with position $\y$ and momentum $\py$, 
described by a Hilbert space $\cK$.
The outcome of the measurement is obtained by measuring the probe position $\y$, 
called the {\em meter observable}, at time $t=\Det$.
The total Hamiltonian for the object and the probe is taken to be
\beql{(1)}
{H}_{\bS+\bP} = {H}_{\bS} + {H}_{\bP} + K{H},
\eeq
where ${H}_{\bS}$ and ${H}_{\bP}$ are the free Hamiltonians
of $\bS$ and $\bP$, respectively, ${H}$ represents the measuring interaction,
and $K$ is the coupling constant.  
We assume that the coupling
is so strong $(K \gg 1)$ 
that ${H}_{\bS}$ and ${H}_{\bP}$ can be neglected.
We choose $\De t$ as $K\De t=1$. 
In the von Neumann model the measuring interaction is given by 
\beql{829o} 
H=\x\otimes\py.
\eeq 
Then, the unitary operator of the time evolution of $\bS+\bP$ from $t=0$ to $t=\ta\le \Det$ is given by
\beql{829p}
U(\ta)=\exp\left(\frac{-iK\ta}{\hbar}\x\otimes\py\right).
\eeq
Suppose that the object $\bS$ and the probe $\bP$ are in the state $\kps$ and $\kxi$, respectively,
just before the measurement; we assume that the wave functions $\ps(x)=\bracket{x|\psi}$ 
and $\xi(y)=\bracket{y|\xi}$ are Schwartz rapidly decreasing functions \cite{RS80},
where $\ket{x}$ and $\ket{y}$ are the position bases of $\bS$ and $\bP$, respectively.

Then, the state of the composite system $\bS+\bP$ just after the measurement is 
$U(\Det)\ket{\ps,\xi}$.
By solving the Schr\"{o}dinger equation,  we have
\beq
\braket{x,y|U(\Det)|\psi,\xi}=\bracket{x|\ps}\bracket{y-x|\xi}.
\eeq

If the observer observes the meter observable $\y(\Det)$ just after the
measuring interaction, the probability distribution of the outcome is given by
\beq
\Pr\{a<\bx\le b \}=\int_{a}^{b}dy \int_{-\infty}^{+\infty} |\braket{x|\psi}|^2\,|\braket{y-x|\xi}|^2\, dx.
\eeq
This shows that if the probe initial wave function $\xi(y)$ approaches to 
the Dirac delta function $\de(x)$, the output probability distribution approaches
to the correct Born probability distribution for the observable $\x$ at time $t=0$.

In the Heisenberg picture, we denote 
\beqas
&\x(\ta)=U(\ta)^{\da}(\x\otimes I) U(\ta),\quad \px(\ta)=U(\ta)^{\da}(\px\otimes I) U(\ta),& \\
&\y(\ta)=U(\ta)^{\da}(I\otimes \y) U(\ta),\quad \py(\ta)=U(\ta)^{\da}(I\otimes \py) U(\ta).&
\eeqas
Solving the Heisenberg equations of motion, 
we have
\beqa
\x(\De t)&=&\x(0),\label{eq:S1}\\
\y(\De t)&=&\x(0)+\y(0),\label{eq:S2}\\
\px(\De t)&=&\px(0)-\py(0),\label{eq:S3}\\
\py(\De t)&=&\py(0).\label{eq:S4}
\eeqa      

\section{Root-mean-square error and disturbance.}
In order to define ``root-mean-square error'' of this measurement,
we recall classical definitions.
Suppose that the true value is given by $\Theta=\theta$ and 
its measured value is given by $\Om=\om$.   
For each pair of values $(\Th,\Om)=(\theta,\om)$, 
the error is defined as $\om-\theta$.  
To define the ``mean error'' with respect to the joint probability distribution 
$ \mu^{\Th,\Om}(d\theta,d\om)$ of $\Th$ and $\Om$, 
Gauss \cite{Gau95} introduced the {\em root-mean-square error} 
$\ve_{G}(\Th,\Om)$ of $\Om$ for $\Th$ as
\beql{rmse}
\ve_{G}(\Th,\Om)=\left(\iint_{\R^{2}}(\om-\theta)^{2} \mu^{\Th,\Om}(\dd\theta,\dd\om)\right)^{1/2},
\eeq
which Gauss  \cite{Gau95} called the ``mean error'' or the ``mean error to be feared,''
and has long been accepted as a standard definition for  the ``mean error.'' 

In the von Neumann model, the value of the observable $\x(0)$
is measured by the value of the meter observable $\y(\De t)$.
Since $\x(0)$ and $\y(\De t)$ commute, as seen from \Eq{S2},
we have the joint 
probability distribution $\mu^{\x(0),\y(\De t)}(\dd\theta,\dd\o)$ of the values of $\x(0)$ and $\y(\De t)$
as
\beql{JPDE}
\mu^{\x(0),\y(\De t)}(\dd\theta,\dd\o)=\bracket{E^{\x(0)}(\dd\theta)E^{\y(\De t)}(\dd\o)},
\eeq
where $E^{A}$ stands for the spectral measure of an observable $A$ \cite{Hal51}, 
and  $\bracket{\cdots}$ stands for the mean value in the state $\ket{\ps,\xi}$.
Then,  from \Eq{rmse} the {\em root-mean-square error} $\ve(\x)$ 
of $\y(\De t)$  for  $\x(0)$ in $\kps$ is given by 
\beqa
\ve(\x)&=&\ve_G(\x(0),\y(\De t))\nn\\
&=&\left(\iint_{\R^{2}}(\o-\theta)^{2}\mu^{\x(0),\y(\De t)}(\dd\theta,\dd\o)\right)^{1/2}\nn\\
&=&\bracket{(\y(\De t)-\x(0))^{2}}^{1/2}\nn\\
&=&\bracket{\y(0)^{2}}^{1/2}.
\label{eq:E-V}
\eeqa

Since $\px(0)$ and $\px(\De t)$ commute, as seen from \Eq{S3},
we have the joint 
probability distribution $\mu^{\px(0),\px(\De t)}(\dd\theta,\dd\o)$ of the values of $\px(0)$ and $\px(\De t)$
as
\beql{JPDD}
\mu^{\px(0),\px(\De t)}(\dd\theta,\dd\o)=\bracket{E^{\px(0)}(\dd\theta)E^{\px(\De t)}(\dd\o)}.
\eeq
The {\em root-mean-square disturbance} $\et(\px)$ of $\px$ from $t=0$ to
$t=\De t$ is defined as the root-mean-square error of $\px(\De t)$ for $\px(0)$ given by 
\beqa
\et(\px)&=&\ve_G(\px(0),\px(\De t))\nn\\
&=&\left(\iint_{\R^{2}}(\o-\theta)^{2}\mu^{\px(0),\px(\De t)}(\dd\theta,\dd\o)\right)^{1/2}\nn\\
&=&\bracket{(\px(\De t)-\px(0))^{2}}^{1/2}\nn\\
&=&\bracket{\py(0)^2}^{1/2}.
\label{eq:D-V}
\eeqa

Then, by the Kennard inequality \eq{Ken27} we have
\beqa
\ve(\x)\et(\px)&=&\bracket{\y(0)^{2}}^{1/2}\bracket{\py(0)^2}^{1/2}\nn\\
&\ge& \si(\y(0))\si(\py(0))\ge\frac{\hbar}{2}.
\eeqa
Thus, the von Neumann model satisfies the Heisenberg EDR \eq{Hei27E}.

Since only the von Neumann model is available as a mathematically solvable mode
of position measurement until 1980's,  model analysis of position measurement 
did not lead to refuting the Heisenberg EDR but rather enforced the belief that good position 
measurements satisfy the Heisenberg EDR \eq{Hei27E}; see for example Refs.~\cite{CTDSZ80,Cav85}.

This belief was also enforced by Arthurs and Kelly \cite{AK65} suggesting that all the joint 
unbiased measurement of position and momentum satisfy the Heisenberg error tradeoff relation.
A {\em joint position-momentum measurement} can be modeled by a triple  $(M_Q,M_P,\kxi)$ consisting 
of commuting meter observables $M_Q$ and $M_P$ in the composite system $\bS+\bP$ described 
by $\cH\otimes\cK$ with the initial state $\kxi\in\cK$ of the probe $\bP$.
Then, the joint measurement $(M_Q,M_P,\kxi)$ is called {\em unbiased} if the mean values 
of meter observables $M_Q$ and $M_P$ coincides with the mean values of $Q$ and $P$, respectively, i.e., 
$\bracket{M_Q}=\bracket{Q}$ and  $\bracket{M_P}=\bracket{P}$, in any input states of $\bS$.
Then, Arthurs and Kelly \cite{AK65} showed that the relation 
\beq
\si(M_Q)\si(M_P)\ge \hbar,
\eeq
holds for any unbiased joint position-momentum measurement $(M_Q,M_P,\kxi)$ in any input state of $\bS$.
It is explained that the lower bound is twice as much as the lower bound for $\si(Q)\si(P)$
because of inevitable errors included in the value of $M_Q$ and $M_P$.
In Ref.~\cite{91QU}, their result was reformulated so that 
the Heisenberg error tradeoff relation 
\beql{ERR}
\ep(Q)\ep(P)\ge\frac{\hbar}{2},
\eeq
holds for any unbiased joint position-momentum measurement $(M_Q,M_P,\kxi)$ in any input state of $\bS$,
where $\ep(Q)$ and $\ep(P)$ are the rms errors of joint measurement of
position $Q$ and momentum $P$ defined through the joint probability distribution of 
$Q$ and $M_Q$ and that of $P$ and $M_P$, which always exist in unbiased case; 
see for another approach Ref.~\cite{Ish91}.
Thus, if we consider unbiased joint measurements as good joint measurements,
we can say that every good joint measurement satisfies the Heisenberg error tradeoff 
relation \eq{ERR}.  

However, this does not imply that every good position measurement satisfies
the Heisenberg EDR \eq{Hei27E}.  Given a position measurement
with the meter observable $M$ in the probe, if the momentum is measured just after the 
position measurement, we have a joint position-momentum measurement 
$(M_Q,M_P,\kxi)$ with $M_Q=M(\De t)$ and $M_P=P(\De t)$.  In this case,
the error tradeoff relation \eq{ERR} is equivalent with the EDR \eq{Hei27E} with $\ep(P)=\et(P)$,  
and we can consider unbiased position measurements as good position
measurements.  Nevertheless, we cannot conclude that every good position measurement
satisfies the Heisenberg EDR \eq{Hei27E}, since an unbiased position measurement
followed by a precise momentum measurement not necessarily
satisfies  $\bracket{P(\De t)}=\bracket{P(0)}$ or $\bracket{M_P}=\bracket{P}$.

\section{Measurement violating the Heisenberg EDR}
In 1980,  Braginsky,  Vorontsov, and Thorne \cite{BVT80} claimed that
the Heisenberg EDR \eq{Hei27E} leads  to 
a sensitivity limit, called the {\em standard quantum limit} (SQL), for 
gravitational wave detectors of interferometer type, which make use of free-mass
position monitoring.  Subsequently,  Yuen \cite{Yue83} questioned the validity of the SQL,
and then Caves \cite{Cav85} defended the SQL by giving a new proof of the SQL
without a direct appeal to \Eq{Hei27E}.
Eventually,  the conflict was reconciled in Refs.~\cite{88MS,89RS} by pointing out that 
Caves \cite{Cav85} used (unfounded) assumption (H1) in his derivation of the SQL,
and a solvable model of an error-free position measurement was constructed
that breaks the SQL (see also Ref.~\cite{Mad88}); later  
this model was shown to break the Heisenberg EDR \eq{Hei27E} \cite{02KB5E}.

In what follows, we introduce this model by
modifying the measuring interaction of the von 
Neumann model. 
In this new model, the object, the probe, and the probe
observables are the same systems and the same observable as the von
Neumann model. The measuring interaction is taken to be \cite{88MS}
\beql{829ox}
H=\frac{\pi}{3\sqrt{3}}
(2\x\otimes\py-2\px\otimes\y
+\x\px\otimes I-I\otimes \y\py).
\eeq
The coupling constant  $K$ and the time duration  $\De t$ are chosen as
before so that $K\gg 1$ and $K\De t=1$.
Then, the unitary operator $U(\ta)$ for $0\le\ta\le\Det$ is given by 
\beql{829px}
U(\ta)=\exp\left[\frac{-i\pi K\ta}{3\sqrt{3}\hbar}
(2\x\otimes \py-2\px\otimes \y
+\x\px\otimes I-I\otimes \y\py)\right].
\eeq

Solving the Heisenberg equations of motion for $t<t+\ta<t+\De t$, we
obtain 
 \beqa
\x(\ta)
 &=&\frac{2}{\sqrt{3}}\x(0)\sin \frac{(1+K\ta)\pi}{3}
 +\frac{-2}{\sqrt{3}}\y(0)\sin \frac{K\ta\pi}{3},\\
\y(\ta)
 &=&\frac{2}{\sqrt{3}}\x(0)\sin \frac{K\ta\pi}{3}
 +\frac{-2}{\sqrt{3}}\y(0)\sin \frac{(1-K\ta)\pi}{3},\\
\px(\ta)
 &=&\frac{-2}{\sqrt{3}}\px(0)\sin \frac{(1-K\ta)\pi}{3}
 +\frac{-2}{\sqrt{3}}\py(0)\sin \frac{K\ta\pi}{3},\\
\py(\ta)
 &=&\frac{2}{\sqrt{3}}\px(0)\sin \frac{K\ta\pi}{3}
 +\frac{2}{\sqrt{3}}\py(0)\sin \frac{(1+K\ta)\pi}{3}.
 \eeqa
For $\ta=\De t=1/K$, we have
\beqa
\x(\De t)&=&\x(0)-\y(0),\label{eq:ozawa-model-1}\\
\y(\De t)&=&\x(0),\label{eq:ozawa-model-2}\\
\px(\De t)&=&-\py(0),\label{eq:ozawa-model-3}\\
\py(\De t)
&=&\px(0)+\py(0)\label{eq:ozawa-model-4}.
\eeqa

As in the von Neumann model, the value of the observable $\x(0)$
is measured by the value of the meter observable $\y(\De t)$.
Since $\x(0)$ and $\y(\De t)$ commute, as seen from \Eq{ozawa-model-2},
we have the joint 
probability distribution $\mu^{\x(0),\y(\De t)}(\dd\theta,\dd\o)$ of the values of $\x(0)$ and $\y(\De t)$ by \Eq{JPDE}
Then,  from \Eq{rmse} the  rms error $\ve(\x)$
of $\y(\De t)$  for  $\x(0)$ in $\kps$ is given by 
\beqa
\ve(\x)
&=&\left(\iint_{\R^{2}}(\o-\theta)^{2}\mu^{\x(0),\y(\De t)}(\dd\theta,\dd\o)\right)^{1/2}\nn\\
&=&\bracket{(\y(\De t)-\x(0))^{2}}^{1/2}\nn\\
&=&0.
\label{eq:E-V-2}
\eeqa

Since $\px(0)$ and $\px(\De t)$ commute, as seen from \Eq{ozawa-model-3},
we have the joint 
probability distribution $\mu^{\px(0),\px(\De t)}(\dd\theta,\dd\o)$ of the values of $\px(0)$ and $\px(\De t)$
by \Eq{JPDD}.
The rms disturbance $\et(\px)$ of $\px$ from $t=0$ to
$t=\De t$ is given by 
\beqa
\et(\px)
&=&\left(\iint_{\R^{2}}(\o-\theta)^{2}\mu^{\px(0),\px(\De t)}(\dd\theta,\dd\o)\right)^{1/2}\nn\\
&=&\bracket{(\px(\De t)-\px(0))^{2}}^{1/2}\nn\\
&=&\bracket{(\py(0)+\px(0))^2}^{1/2}<\infty.
\label{eq:D-V}
\eeqa

Consequently,  we have 
\beq
\ep(\x)\et(\px)=0.
\eeq
Therefore, our model obviously violates the Heisenberg EDR \eq{Hei27E}.  

Taking advantage of the above model, the argument was refuted
that the uncertainty principle generally leads to the SQL claimed 
in Ref.~\cite{BVT80} for monitoring free-mass position \cite{Yue83,88MS}. 

If $\bracket{\px(0)^{2}}\to 0$
and $\bracket{\py(0)^{2}}\to 0$ (i.e.,  $\kps$ and $\kxi$
tend to the momentum eigenstate with zero momentum) then
we have even $\et(\px(t))\to 0$ with $\ep(\x)=0$.
Thus, we can measure position precisely without effectively disturbing 
momentum in a near momentum eigenstate;
see Ref.~\cite{01CQSR} for detailed discussion on the 
quantum state reduction caused by the above model. 

As shown above,  the Heisenberg EDR \eq{Hei27E} is taken to be a breakable 
limit \cite{GLM04}, but then the problem remains:
what is the unbreakable constraint between error and disturbance,
which Heisenberg originally intended?  

\section{Universally valid EDR}
In 2003,  
the present author \cite{03HUR,03UVR,03UPQ} showed the  relation
\beql{UEDR}
\ep(A) \et(B)+|\bracket{[n(A),B]}+\bracket{[A,d(B)]}|
\ge  \frac{1}{2}\left| \langle  [A,B]  \rangle \right|,
\eeq
which is universally valid for any observables $A,B$, any system state, and 
any  measuring apparatus,
where $n(A)$ and $d(B)$ are system observables representing the first moments of 
the error and the disturbance
for $A$ and $B$, respectively.
From \Eq{UEDR}, it is concluded that if the error and the disturbance are statistically 
independent from system state,  then the Heisenberg EDR
\beql{HEDR}
\ep(A)\et(B)\geq\frac{1}{2}|\bracket{[A,B]}|
\eeq
holds, extending the previous results \cite{AG88,Ray94,91QU,Ish91}.
The additional correlation term  in \Eq{UEDR} allows the error-disturbance product 
$\epsilon (A) \eta (B)$ to violate the Heisenberg EDR \eq{HEDR}.
In general,  the relation
\beql{OEDR}
\epsilon (A) \eta(B)+\epsilon(A)\sigma(B)+\sigma(A)\eta(B) 
\ge  \frac{1}{2}\left| \langle  [A,B]  \rangle \right|,
\eeq
holds for any observables $A,B$, any system state, and 
any measuring apparatus \cite{03HUR,03UVR,03UPQ,04URJ,04URN,05UUP}.

The new relation
\eq{OEDR} leads to the following new constraints for error-free 
measurements and non-disturbing measurements: if $\ep(A)=0$ then
\beql{DB}
\sigma(A)\eta(B) \ge  \frac{1}{2}\left| \langle  [A,B]  \rangle \right|,
\eeq
and if $\et(B)=0$ then
\beql{EB}
\epsilon(A)\sigma(B)\ge  \frac{1}{2}\left| \langle  [A,B]  \rangle \right|.
\eeq
Note that if $\bracket{[A,B]}\not=0$,
Heisenberg EDR \eq{HEDR} leads to divergences in both cases.
The new error bound \Eq{EB} was used to derive a conservation-law-induced limits for
measurements \cite{03UPQ,04UUP} (see also \cite{02CLU,BL11})
quantitatively generalizing the 
Wigner-Araki-Yanase theorem \cite{Wig52,AY60,Yan61,91CP}
and was used to derive an accuracy limit for quantum computing induced 
by conservation laws \cite{03UPQ} 
(see also \cite{02CQC,Ban02b,03QLM,05CQL,06MEP,07CLI,09GFA}).

\section{Operator formalism for error and disturbance}
To derive the above relations, consider a measuring process $\bM=(\cK,\kxi,U,M)$
determined by the probe system $\bP$ described by a Hilbert space $\cK$, 
the initial probe state $\kxi$, 
the unitary evolution $U$ of the composite system $\bS+\bP$ during the measuring interaction,
and the meter observable $M$ of the probe $\bP$  to be directly observed \cite{03UVR}.
We assume that the measuring interaction turns on at time $t=0$ and turns off at time $t=\De t$.
In the Heisenberg picture, we write
$$
A_1(0)=A_1\otimes I, \quad A_2(0)=I\otimes A_2, \quad A_{12}(\Det)=U^{\da}A_{12}(0)U,
$$
for an observable $A_1$ of $\bS$, an observable $A_2$ of $\bP$, and an observable $A_{12}(0)$ of $\bS+\bP$.

The {\em error observable} $N(A)$ representing the difference between the measured observable $A(0)$
and the meter observable $M(\Det)$ to be read and 
the {\em disturbance observable} $D(A)$ representing the change 
in $B$ caused by the measuring interaction are defined by
\beqa
N(A)&=&M(\Det)- A(0),\\
D(B)&=&B(\Det)-B(0).
\eeqa
The the {\em mean error operator} $n(A)$ and the {\em mean disturbance operator} $d(B)$ in \Eq{UEDR} are defined by
\beqa
n(A)&=&\bracket{\xi|N(A)|\xi},\\
d(B)&=&\bracket{\xi|D(B)|\xi}.
\eeqa
The {\em (root-mean-square) error} $\ep(A,\rh)$ 
and the {\em (root-mean-square) disturbance} $\et(B,\rh)$ for observables $A,B$ and state (density operator) $\rh$
on a Hilbert space $\cH$ were defined by
\beqa
\ep(A,\rh)^2&=&\Tr[N(A)^2\rho\otimes \ketbra{\xi}], \\
\et(B,\rh)^2&=&\Tr[D(B)^2\rho\otimes \ketbra{\xi}].
\eeqa

The definition of $\ep(A)=\ep(A,\rh)$ is uniquely derived from the classical notion of root-mean-square 
error if $M(\De  t)$ and $A(0)$ commute \cite{13DHE},  as in the models discussed in the previous sections.
Otherwise, it is considered as a natural quantization of the notion of classical 
root-mean-square error.
It is also pointed out that $\ep(A)$ coincides with the root-mean-square error
of a quantum estimator for an orthogonal pure state estimation problem with the 
uniform prior distribution \cite{14NDQ}.
The definition of $\eta(B)=\eta(B,\rh)$ is derived analogously,  although there are recent debates 
on alternative approaches \cite{BHL04,BLW13, RMHS13,13DHE}.

In particular, Busch, Heinonen, and Lahti \cite{BHL04} pointed out that there is a case
in which $\ep(A)=0$ holds but  $A$ cannot be considered to be measured precisely.
In response to this, we characterized the case where $A$ is measured precisely as follows  \cite{05PCN,06QPC}.

We say that the measuring process $\bM$ {\em precisely measures} an observable $A$ in a state $\rh$ 
if observables $A(0)$ and $M(\De t)$ commute in the state $\rh\otimes\ketbra{\xi}$
and the joint probability distribution $\mu^{A(0),M(\De t)}$
of  $A(0)$ and $M(\De t)$ concentrates on the diagonal, i.e.,  
\beq
\mu^{A(0),M(\De t)}(\{(x,y)\in\R^{2}\mid x\not= y\}=0.
\eeq
The {\em weak joint distribution} $\mu_{W}^{A(0),M(\De t)}$ of $A(0)$ and $M(\De t)$ in a state $\rh$
is defined  by
\beq
\mu_{W}^{A(0),M(\De t)}(dx,dy)=\bracket{E^{A(0)}(dx)E^{M(\De t)}(dy)}.
\eeq
The joint probability distribution $\mu^{A(0),M(\De t)}$ exists only when $A(0)$ and $M(\De t)$ commute
in the state $\rh\otimes\ketbra{\xi}$, while the weak joint distribution $\mu_{W}^{A(0),M(\De t)}$ always exists.
The {\em cyclic subspace} $\cC(A,\rh)$ generated by $A$ and $\rh$ is defined  as the closed subspace of $\cH$
generated by $\{E^{A}(\De)\kph\mid \De\in\cB(\R), \kph\in\rh\cH\}$, where $\cB(\R)$ is the Borel $\si$-field 
of the real line $\R$.
A {\em generating subset} of $\cC(A,\rh)$ is a set $\cS$ of vector states $\kph\in\cH$ such that 
$\cS^{\perp\perp}=\cC(A,\kph)$, where $\perp$ stands for the orthogonal complement.
Then, the following theorem holds \cite{05PCN,06QPC}.

\begin{Theorem}
Let $\bM=(\cK,\kxi,U,M)$ be a measuring process for the system $\bS$ described by a Hilbert space $\cH$.
Let $A$ be an observable of $\bS$ and $\rh$ be a state of $\bS$. 
Then, the following conditions are equivalent.
\begin{enumerate}
\item The measuring process $\bM$ precisely measures observable $A$ in state $\rh$.
\item  The weak joint distribution $\mu_{W}^{A(0),M(\De t)}$ in state $\rh$ concentrates on the diagonal, i.e., 
$$
\bracket{E^{A(0)}(\De)E^{M(\De t)}(\Ga)}=0
$$
if $\De\cap\Ga=\emptyset$.
\item $\ep(A,\kph)=0$ for all $\kph\in\cC(A,\rh)$.
\item There exists a generating subset $\cS$ of $\cC(A,\rh)$ such that $\ep(A,\kph)=0$ for all $\kph\in S$.
\end{enumerate}
\end{Theorem} 

We say that the measuring process $\bM$ {\em does not disturb} an observable $B$ in a state $\rh$ 
if observables $B(0)$ and $B(\De t)$ commute in the state $\rh\otimes\ketbra{\xi}$ 
and the joint probability distribution $\mu^{B(0),B(\De t)}$
of  $B(0)$ and $B(\De t)$ concentrates on the diagonal.
The non-disturbing measuring processes defined above can be characterized analogously.
  
From the above theorem, we can conclude that $\ep(A)$ is negatively biased in the
sense that positive error $\ep(A)>0$ always implies that the $A$ cannot be measured precisely.
Thus, a non-zero lower bound for $\ep(A)$ indicates a limitation for precise measurements.
For $\et(B)>0$ we have an analogous conclusion.
Moreover,  the above characterizations of precise
and non-disturbing measurements lead to the following 
definitions of the {\em locally uniform root-mean-squre error} $\overline{\ep}(A,\rh)$
and the {\em locally uniform root-mean-squre disturbance} $\overline{\et}(B,\rh)$
\cite{06NDQ}: 
\beqa
\overline{\ep}(A,\rh)&=&\sup_{\kph\in\cS}\ep(A,\kph),\\
\overline{\et}(B,\rh)&=&\sup_{\kph\in\cS}\et(B,\kph),
\eeqa
where $\cS=\cC(A,\rh)$ or $\cS$ is a generating subset of 
$\cC(A,\rh)$.
Then, it is shown that 
$\overline{\ep}(A,\rh)=0$ if and only if the measurement precisely
measures $A$ in $\rh$, and that $\overline{\et}(B,\rh)=0$ if and only if the measurement does not disturb $B$ in $\rh$.  
For those quantities, Heisenberg's EDR 
\beq
\overline{\ep}(Q,\rh)\overline{\et}(P,\rh)\ge\frac{\hbar}{2}
\eeq
is still violated by a linear position measurement \cite{06NDQ}, and 
the relation
\beqa
\overline{\ep}(A)\overline{\et}(B)+\overline{\ep}(A)\sigma(B)
+\sigma(A)\overline{\et}(B)\geq\frac{1}{2}|
	\bracket{[A,B]}|
	\label{OZA}\nn\\
\eeqa
holds universally \cite{06NDQ}, where
$\overline{\ep}(A)=\overline{\ep}(A,\rh)$ and
$\overline{\et}(B)=\overline{\et}(B,\rh)$.

\section{Experimental tests}
There has been a controversy \cite{Wer04,KS05} on the question 
as to whether the rms error and rms disturbance are experimentally 
accessible without knowing the details of 
measuring process $(\cK,\kxi,U,M)$ and the state $\rh$.
To clear this question two methods have been proposed so far: the
``three-state method''  proposed by the present auhor~\cite{04URN}
and  the ``weak-measurement method'' proposed by Lund-Wiseman~\cite{LW10} 
based on the relation between the rms error/distrubance and the weak joint distribution \cite{88MS,91QU,05PCN}.
The three-state method was demonstrated for qubit systems:
projective measurement of a neutron-spin qubit \cite{12EDU,13VHE} and generalized 
measurement of a photon-polarization qubit \cite{13EVR}. 
The weak-measurement method was demonstrated for generalized measurement 
of a photon-polarization qubit by Rozema {\it et al} \cite{RDMHSS12}, Baek {\it et al} \cite{14ETE}, 
and Ringbauer {\it et al.} \cite{RBBFBW14}.
All the above experiments observed that the Heisenberg EDR \eq{HEDR}
does not hold, while the universally valid EDR \eq{OEDR} and a new stronger universally valid
EDR recently proposed by Branciard \cite{Bra13} hold.
Very recently, we have proposed the third method, the ``two-point quantum correlator method,''
for measuring rms error and disturbance with an experimental proposal for quit measurements  
in Ref.~\cite{14DOO}.  

\ack
This work was supported by MIC SCOPE, No.~121806010, and
the John Templeton Foundation, ID~\#35771.

\section*{References}
%\bibliographystyle{iopart-num}
%\bibliography{myelist,myebib}
\providecommand{\newblock}{}

\end{document}